%% file: vgcompressive.tex
\documentclass[useAMS,usenatbib]{mnras}

\def\draftversion{1} % 0 = clean     1 = draft       2 = referee (clean but with \new commands in bold)
\input{header.tex}

\defcitealias{Agertz2021}{Paper I}
\defcitealias{Renaud2021}{Paper II}
\defcitealias{Renaud2021b}{Paper III}
\defcitealias{Segovia2022}{Paper IV}

\newcommand{\paperi}{\citetalias{Agertz2021}\xspace}
\newcommand{\paperii}{\citetalias{Renaud2021}\xspace}
\newcommand{\paperiii}{\citetalias{Renaud2021b}\xspace}
\newcommand{\paperiv}{\citetalias{Segovia2022}\xspace}

\title[The merger-starburst connection]{The merger-starburst connection across cosmic times}
\author[Renaud et al.] {Florent~Renaud\thanks{florent@astro.lu.se}, \'Alvaro~Segovia~Otero \& Oscar~Agertz\\
\lund}

\date{Accepted 2022 September 6. Received 2022 August 21; in original form 2022 July 20}

\begin{document}
\maketitle

%%%%%%%%%%%%%%%%%%%%%%%%%%%%%%%%%%%%%%%%%%%%%%%%%%%%%%%%%%%%%%%%%%%%%%%%%%%%%%%%%

\begin{abstract}
The correspondence between galaxy major mergers and starburst activity is well-established observationally and in simulations of low redshift galaxies. However, the evolution of the properties of interactions and of the galaxies involved suggests that the starburst response of galaxies to merger events could vary across cosmic time. Using the \vintergatan cosmological zoom-in simulation of a Milky Way-like galaxy, we show here that starbursts, i.e. episodes of fast star formation, are connected with the onset of tidal compression, itself induced by mergers. However, this compression becomes strong enough to trigger starbursts only after the formation of the galactic disc. As a consequence, starburst episodes are only found during a precise phase of galaxy evolution, after the formation of the disc and until the last major merger. As the depletion time quantifies the instantaneous star formation activity, while the specific star formation rate involves the integrated result of the past activity (via the stellar mass), starburst episodes do not necessarily coincide with elevated specific star formation rate. This suggests that not all starburst galaxies are outliers above the main sequence of galaxy formation.
\end{abstract}

%%%%%%%%%%%%%%%%%%%%%%%%%%%%%%%%%%%%%%%%%%%%%%%%%%%%%%%%%%%%%%%%%%%%%%%%%%%%%%%%%
\section{Introduction}

Enhancements of star formation and starbursts are often associated with interactions and mergers, including at early stages of interactions and at high galactic separations \citep[e.g.][]{Armus1987,Kennicutt1987, Xu1991, Sanders1996,Ellison2008,Scudder2012,Ellison2013, Knapen2015, xu2021, Silva2021, Horstman2021}. The effects of individual mergers on the star formation activity have been studied in depth with dedicated simulations of pairs of galaxies in isolation \citep[e.g.][]{Hernquist1989, Barnes1991, Springel2005, Dimatteo2008, Saitoh2009, Teyssier2010, Karl2010, Moreno2015, Renaud2014b, Renaud2015, Renaud2016, Renaud2018, Moreno2019, Linden2022, Petersson2022}. During interactions, torque-driven nuclear inflows, shocks, and tidal compression increase the amount of star forming gas, which in turns boosts the star formation rate (SFR) over timescales of $\sim 10\mh 100 \myr$ \citep{Keel1985, Barnes1991, Jog1992, Renaud2014b}. This, in addition to fueling and quenching mechanisms, demonstrates that the environment of a galaxy is of paramount importance to drive its star formation history. 

Most of the pioneer works on mergers and the associated enhancement of star formation studied low redshift galaxies. However, in the last decade, surveys have targeted the question of the efficiency of high redshift mergers ($z\gtrsim 0.5\mh 1$) at triggering starbursts. Let us focus here on two main results. (\emph{i}) The fraction of outliers above the galaxy main sequence\footnote{i.e. galaxies with a higher specific SFR (sSFR, that is the SFR normalised by the stellar mass) than most galaxies at the same redshift \citep{Elbaz2007, Noeske2007, Rodighiero2011}.} increases much slower with redshift than the merger rate (up to at least $z \approx 4$, \citealt{Schreiber2015, Pearson2019}). (\emph{ii}) The depletion time (i.e. the gas mass divided by the SFR) is significantly shorter at $z \gtrsim 0.5$ than at lower redshift (as shown for instance in the PHIBSS survey, \citealt{Tacconi2018})\footnote{The observational measurements of the depletion time suffer from uncertainties on the CO luminosity to molecular gas mass conversion factor \aco. While observations of local ultra-luminous infrared galaxies \citep{Downes1998} and simulations of major mergers \citep{Renaud2019} both advocate for a lower \aco in local mergers than the Galactic value, varying \aco might be less important at high redshift (see \citealt{Genzel2015, Scoville2016} for a comparison between CO- and dust-based estimates).}. The first point suggests that \emph{if} outliers above the main sequence are starbursts, high redshift mergers are statistically inefficient at triggering starbursts \citep{Jogee2009, Kaviraj2013}. The second implies that \emph{if} mergers are the main (or even unique) source of short depletion times, frequent mergers induce a large number of starburst episodes at high redshift. This contradiction can be resolved in two ways. The first option is that there is no direct correspondence between starbursts and outliers above the main sequence (at high redshift). In other words, the definitions of starburst as a galaxy with a short depletion time (as commonly used in the literature, see e.g. \citealt{Daddi2010, Genzel2010, Kennicutt2021}) and an outlier above the main sequence (e.g. \citealt{Elbaz2007, Sargent2014, Ellison2020, Wang2022}) are not equivalent. The other possibility is that starbursts can be caused by other mechanisms than mergers.

To date, simulations have still not reached a consensus in identifying under which conditions mergers induce starbursts or not, specially at high redshift. For instance, \citet{Perret2014} found that major mergers of gas-rich galaxies (mimicking discs at $z\approx 1\mh 2$, but run without cosmological context) do not trigger strong enhancements of the SFR \citep[see also][]{Scudder2015,Fensch2017}. The high gas fractions of these galaxies have been proposed as the cause of this inefficiency, possibly because of an intrinsically high level of turbulence and its saturation during interactions \citep{Fensch2017}. Yet, the gas-rich mergers of \citet{Moreno2021} do boost the SFR by factors of several 10s, thus questioning the role of the gas fraction on the triggering of starbursts.

In the last years, the influence of mergers on the star formation history of galaxies has started to be studied using cosmological simulations \citep[e.g.][]{Patton2020, Blumenthal2020, Renaud2021, Quai2021, Li2022, Sparre2022}. Several potentially key aspects of this topic are indeed only accessible in cosmological context, like the cosmic evolution of the orbital properties of the interactions (e.g. shorter timescales at high $z$), the effect of repeated mergers, the accretion of intergalactic gas, and the intrinsic evolution of the galaxies. However, star formation is notoriously sensitive to numerical resolution, in particular during starburst episodes \citep{Teyssier2010, Renaud2014b}, which could hamper results from resolution-limited large-scale cosmological volumes. When examining the star forming galaxies in the Illustris simulation, \citet{Sparre2015} noted a significantly lower number of outliers above the main sequence than observed, and interpreted it as a deficit of starbursts in the simulation. Conversely, the \vintergatan cosmological zoom-in simulation does retrieve the short depletion times observed at $z \gtrsim 1$ (see \citealt{Segovia2022}, and the rest of the present paper). Differences in resolution and in the implementation of sub-grid physics still hinder the field and have prevented reaching definite conclusions.
 
In this paper, we address the question of the cosmic evolution of the interaction-driven trigger of starbursts, and of the response of a Milky Way-like galaxy. We use a cosmological zoom-in simulation with a resolution ($20 \pc$) of the same order as that commonly reached in non-cosmological runs ($\sim 1 \mh 50 \pc$). With this setup, we capture jointly the intrinsic evolution of a Milky Way-like disc galaxy, and that of its environment. We particularly focus on the connection between the merger activity, tidal effects, and accelerated star formation, across several phases of the star formation history.

%%%%%%%%%%%%%%%%%%%%%%%%%%%%%%%%%%%%%%%%%%%%%%%
\section{Background and simulation}

In this section, we summarize the key concepts and earlier results needed before presenting the conclusions of this paper.

%%%%%%%%%%%%
\subsection{Tidal field and compression-induced starbursts}

In interactions and mergers, tides play a central role in the morphological transformation of the progenitors, but also in the structure and properties of the interstellar medium. Any tidal field can be described mathematically as minus the Hessian matrix of the gravitational potential, the so-called tidal tensor. Following the formalism introduced in \citet{Renaud2011}, at any position in space, the components of the tidal tensor read
\begin{equation}
T^{ij} = -\frac{\partial^2 \phi}{\partial x^i \partial x^j},
\end{equation}
where $\phi$ is the potential at this position, and $x^i$ is the spatial coordinate along the $i$-th dimension. This $3\times 3$ tensor is real and symmetric, and thus can be written in diagonal form. After this transformation, the three eigenvalues $\{\lambda_1, \lambda_2, \lambda_3\}$ represent the strength of the tidal force along the corresponding eigenvectors. By convention, we order the eigenvalues: $\lambda_1 \ge \lambda_2 \ge \lambda_3$. In the case of all three eigenvalues being negative (i.e. $\lambda_1 < 0$), the tidal field is compressive, meaning that the tidal forces points inward along all directions\footnote{contrarily to e.g. the textbook case of the Earth under the tidal influence of the Moon, where the tidal forces along the Earth-Moon axis point outward.}. One can show that, in classical potentials yielding tidal compression, the ratio of the eigenvalues is typically of the order of unity (\citealt{Renaud2017}). Therefore, and for simplicity, we limit the analysis in this paper to that of the main eigenvalue $\lambda_1$. 

Analytical models and simulations show that, in any galaxy interaction, the superposition of two galactic potentials induces the formation of cores in the gravitational potential, possibly over scales of several kpc \citep{Renaud2008,Renaud2009}. The inflection of the potential in cores marks the boundary of tidal compression. This tidal compression is maximal during pericenter passages, when the overlap of the potentials encompasses the largest amounts of galactic matter. \citet{Renaud2014b} found that the onset of tidal compression is shortly followed by an increase of the velocity dispersion of the gas in these regions. This enhanced dispersion is of different nature from the classical turbulence: its compressive mode dominates over the solenoidal (mixing) one, contrarily to what is found in classical interstellar turbulence where the two modes are in energy equipartition \citep[e.g.][]{Kritsuk2007, Federrath2010}. The tidal and turbulent compression of the gas from the large galactic scales down to the small cloud scales induces an excess of dense gas (but not necessarily an increase of the maximum density), which in turn triggers a starburst event in the form of a drop of the depletion time. Tidal compression and direct shocks of the gas reservoirs \citep{Jog1992} explain off-centered starbursts in interacting galaxies, but only tidal compression can explain them during non-penetrating encounters (i.e. during the pre-coalescence phases), and in regions where the gas reservoirs of the two galaxies do not overlap. 

The triggering of starbursts by tidal and turbulent compression has been found in simulations of several distinct merger configurations \citep[e.g.][]{Renaud2014b, Renaud2016, Renaud2018, Fensch2017}, and using different numerical methods \citep[e.g.][]{Sparre2022, Li2022}. However, the short dynamical timescales in such systems make it difficult to spatially associate a given compression volume with a starburst event occurring a few $10 \Myr$ later, when the initial trigger has long left the area \citep{Renaud2019b}.

%%%%%%%%%%%%
\subsection{The \vintergatan simulation}
\label{sec:vg}

In this paper, we follow the cosmological evolution of a Milky Way-like galaxy using the \vintergatan simulation (\citealt{Agertz2021, Renaud2021, Renaud2021b}, hereafter \paperi, \paperii, \paperiii, respectively). \vintergatan is a cosmological zoom-in simulation of a Milky Way-mass galaxy\footnote{Simulation movies are available here:\\\url{https://www.astro.lu.se/~florent/vintergatan.php}}, down to a resolution of $20 \pc$. The time sampling of the snapshots used in the post-process analysis is of the order of $100 \mh 150 \Myr$. The simulation is run with the adaptive mesh refinement code \ramses \citep{Teyssier2002} and includes treatments of heating, cooling, star formation, and stellar feedback \citep{Agertz2013, Agertz2015}, as detailed in \paperi. 

In this paper, the tidal field is computed using the first-order finite differences of the gravitational acceleration over a scale of $\approx 200 \pc$. We have performed the same calculations at scales 4 and 8 times higher and lower. This change quantitatively affects the intensity of the tidal accelerations by factors of a few, but not qualitatively the trends and conclusions presented below.

The analysis of the evolution of the star formation activity of \vintergatan reveals 3 distinct phases (see \citealt{Segovia2022}, hereafter \paperiv, for details):
\begin{itemize}
\item \emph{Early} phase ($z>4.8$): before the disc is in place, the numerous mergers do not trigger starburst episodes, and the depletion time is long ($\sim 1 \Gyr$).
\item \emph{Starburst} phase ($1.0 < z < 4.8$): repeated starbursts induce drops of the depletion time ($\sim 100 \Myr$).
\item \emph{Secular} phase ($z < 1.0$): after the last major merger, the galaxy returns to long depletion times ($\sim 1 \Gyr$).
\end{itemize}
The events associated with each phase and the transitions between them are generic features in the evolution of a disc galaxy (disc formation, major mergers, end of the merger phase). Therefore, we suggest that these phases can be transposed to the evolution of any disc galaxy (with a possible adjustment of the timing).

%%%%%%%%%%%%%%%%%%%%%%%%%%%%%%%%%%%%%%%%%%%%%%%
\section{Results}

%%%%%%%%%%%%%%%%
\subsection{Intensity of tidal compression}
\label{sec:intensity}

\begin{figure}
\centering
\includegraphics{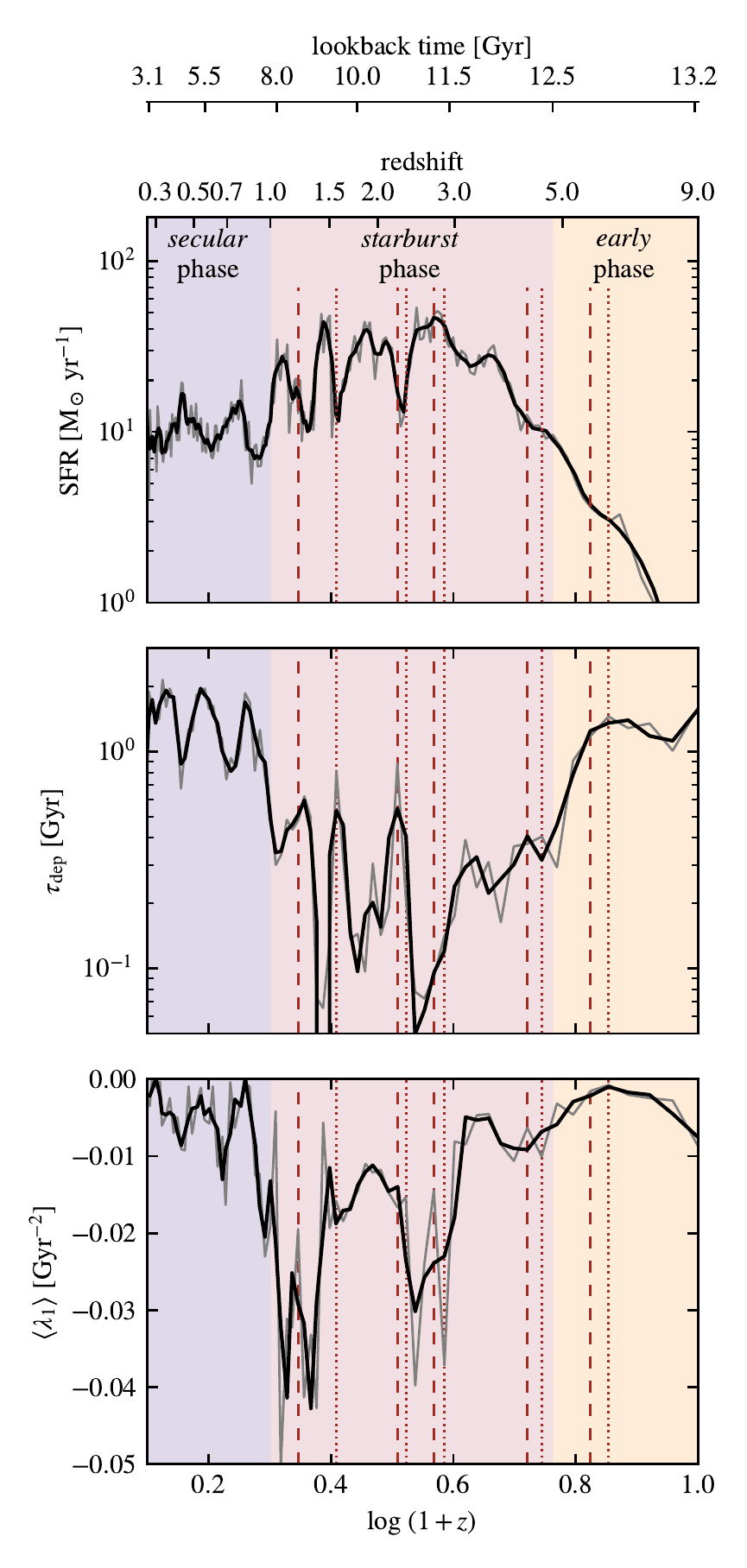}
\caption{The average strength of the compressive tidal field (bottom panel, represented here by the mass-weighted average of the main eigenvalue of the tidal tensor) increases, i.e. becomes more negative, near the peaks of the star formation rate (top panel) and the drops of the depletion times (middle panel). Only regions in compressive tides ($\lambda_1 < 0$) are considered. The logarithmic scaling of the abscissa axis is chosen for readability. The raw data (grey) is smoothed (black) using the Savitzky-Golay algorithm to mitigate the effect of time discreteness. Vertical lines mark the first pericenter passage (dotted) and the start of the final coalescence (dashed) for each major mergers (mass ratio $>$ 1:10, see \paperii). The three shaded areas indicate the phases identified in \paperiv (recall \sect{vg}).}
\label{fig:avel1}
\end{figure}

To assess the importance of tidal compression on the star formation activity, one needs to consider two aspects: the intensity of the compression, and the amount of gas in compressive volumes. \fig{avel1} illustrates the former. We compare the evolution of the strength of the tidal field to that of the SFR and the depletion time (as already introduced in \paperi, \paperii and \paperiv), and highlight the synchronism of these 3 quantities. The intensity of compression is evaluated as the mass-weighted average of the main eigenvalue ($\lambda_1$) of the tidal tensor, only where it takes negative values (i.e. in compressive tides) and within 3 times the stellar half-mass radius of \vintergatan.

The \emph{early} ($z> 4.8$) and \emph{secular} ($z < 1.0$) phases have low SFRs ($\lesssim 10 \msunyr$) and long depletion times ($\sim 1 \Gyr$), indicating respectively a weak and slow star formation activity. While this is expected in the \emph{secular} phase of evolution for a galaxy on the main sequence \citep{Speagle2014}, it seems counter-intuitive during the galactic bombardment by numerous mergers in the \emph{early} phase (as expected from the observed and simulated increase of the merger rate with redshift, see \citealt{Fakhouri2010, Schreiber2015} among many others). In \paperiv, we showed that the lack of starburst response of the galaxy to the merger stimuli at high redshift ($z \gtrsim 4.8$) is linked with the absence of ordered, rotation-supported dynamics, before the formation of the disc. Then, the velocity dispersion of the gas remains at high levels but is not dominated by the compressive mode of turbulence. This turbulent support prevents the rapid formation and collapse of clouds, and thus the star formation activity is not bursty. 

In terms of tides, the \emph{early} and \emph{secular} phases yield relatively weak compression, with limited variations in amplitude. The dips seen in the \emph{secular} phase are associated with the few minor mergers pacing the late evolution (\paperii). They do not have counterparts in the \emph{early} phase, but this absence could be due to the low time sampling of the simulation, compared to the short dynamical time at this epoch.

Conversely, during the \emph{starburst phase} ($1.0 < z < 4.8$), every major interaction (i.e. pericenter passages and coalescences) lead to rapid enhancements of the SFR, and drops of the depletion time. Reduced depletion times at intermediate redshifts ($z \sim 0.6\mh 4$) are also been reported statistically from observational surveys of star forming galaxies (e.g. PHIBSS, \citealt{Tacconi2018}). In \paperiv, we showed that the onset of the \emph{starburst} phase corresponds to the epoch when rotational motions overtake dispersion, i.e. when the galactic disc starts to form. Then, the galactic matter yields an organized, well-ordered, and collective response to tidal interactions. Here, we link this response with the apparition of volumes of strong tidal compression (i.e. very negative $<\lambda_1>$) associated with drops of the depletion time.

Overall, during all the phases of star formation, we note an excellent agreement between the depletion time and the intensity of tidal compression, even at high redshift.

%%%%%%%%%%%%%%%%
\subsection{Masses and locations in compressive tides}

\begin{figure}
\centering
\includegraphics{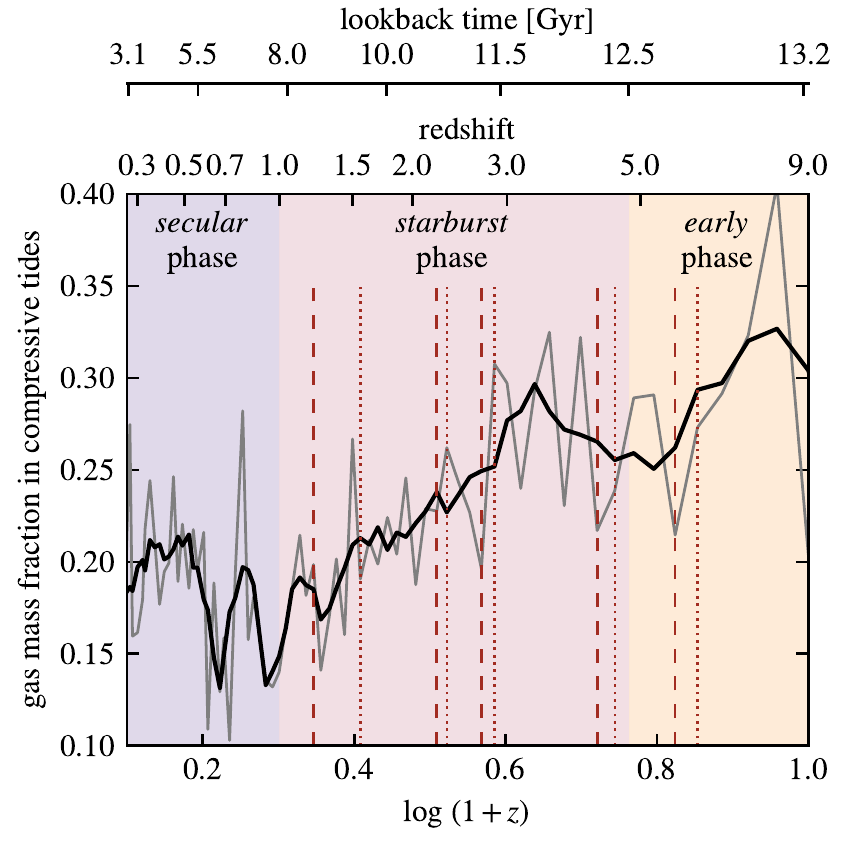}
\caption{Evolution of the gas mass fraction in compressive tides. Only the gas within three stellar half-mass radii is considered. As in \fig{avel1}, the raw data (grey) is smoothed using the Savitzky-Golay algorithm (black line). Although major mergers seem to punctually increase (or slow down the decrease) of this fraction, except in the \emph{early} phase, the overall evolution is a decrease during the \emph{starburst} phase, between $z \approx 3$ and $z \approx 1$.}
\label{fig:compressivemassfraction}
\end{figure}

In \fig{compressivemassfraction}, we quantify the fraction of gas in compressive tides ($\lambda_1 < 0$), within 3 times the stellar half-mass radius of \vintergatan. Tidal compression concerns about one third of the gas mass at high redshift ($z \gtrsim 3$), but this fraction decreases during the \emph{starburst} phase to $15\mh 20 \%$. Major mergers only have a mild and short-lived influence on this trend: pericenter passages and coalescences either slightly increase or slow down the decrease of this fraction. We note however that the time resolution of the snapshots of this simulation ($\approx 150 \Myr$) is significantly longer that the timescale for tidal compression ($\sim 10\mh 50 \Myr$, \citealt{Renaud2009}). Therefore, it is possible that the full extent of compressive events (rise, peak, decline) is not always properly captured by our time sampling.

\begin{figure}
\centering
\includegraphics{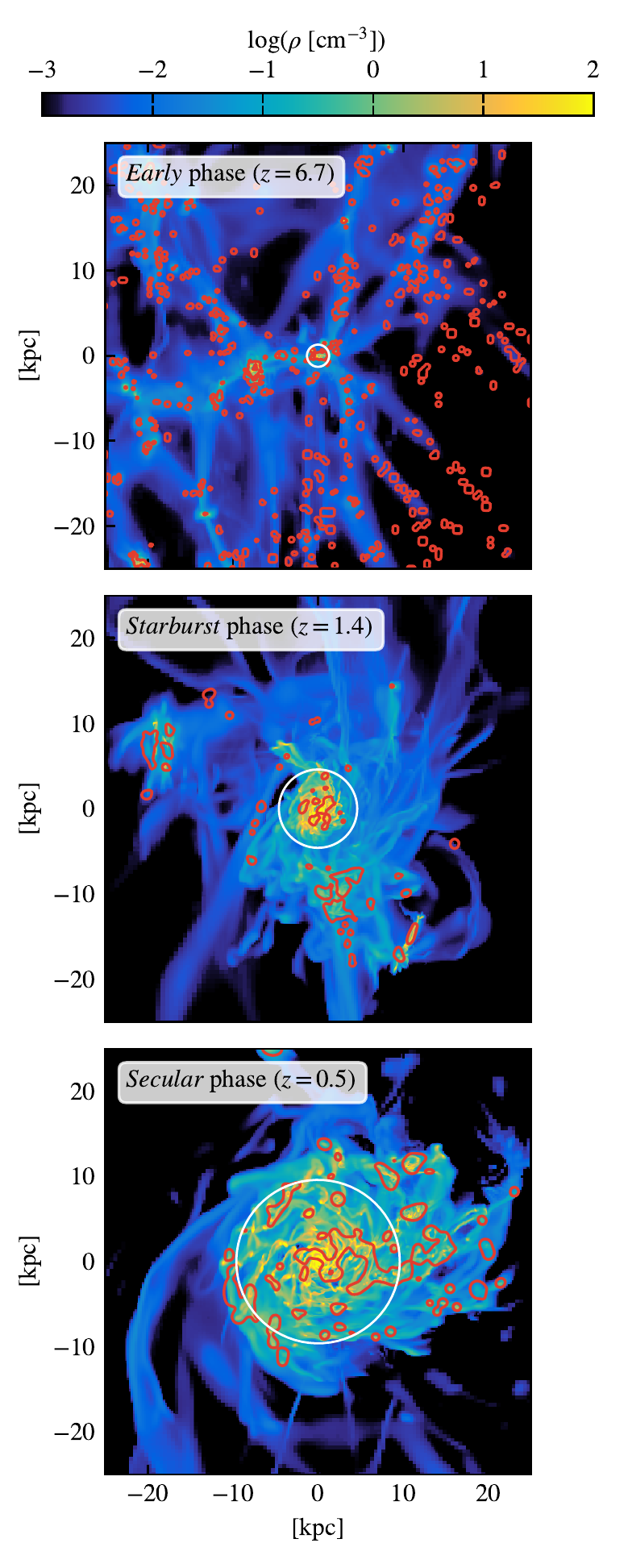}
\caption{Maps of the gas density centered on the main galaxy, from three snapshots illustrative of the three evolutionary phases. For these selected snapshots, 3 times the stellar half-mass radius (which we use here as a proxy for the size of the galaxy) corresponds to 1.3, 4.6, and $9.6 \kpc$ and is shown by the white circles. The contours indicates the regions of tidal compression ($\lambda_1 < 0$) at the point along the projected dimension where the gas density is the highest.}
\label{fig:maps}
\end{figure}

In the \emph{early} phase, the moderate increase of the mass fraction in compressive tides caused by the only major merger is mitigated by the low intensity of the compression at this epoch (\fig{avel1}). This further shows that mergers events are not efficient stimuli for compression in young galaxies before the formation of their disc. However, the overall mass fraction in compressive tides is at its highest level in this phase. This is caused by the presence of a large number of galaxies in a small volume due to the compactness of the Universe at high $z$. The multitude of overlapping potentials creates a large volume of tidal compression in and around the young galaxies. This is illustrated in \fig{maps} by the contours marking the positions of compressive tides on gas density maps.

As young galaxies begin their assembly, their potential is still shallow. This allows for marginally compressive regions ($\lambda_1$ close to 0) over large volumes. The strongest compression is found in-between the galaxies (caused by the overlap of their potential, and probably also by the density profiles of cosmological filaments, \citealt{Zhu2021}). However, in the low densities of the intergalactic medium, this concerns only a small fraction of the gas mass. It is possible that such an intergalactic compression could favor the formation of new gas structures, and even lead to the formation of stars and star clusters. The small scales involved and the technical challenges of properly capturing the thermodynamics of low density circumgalactic and intergalactic media \citep{McCourt2012, Tumlinson2017} prevent us from drawing firm conclusions on this particular point.

During the \emph{starburst} phase, the potential of the galaxy allows for stronger compression (recall \fig{avel1}), but mostly in the outer galaxy and the intergalactic medium, while the inner, dense galaxy develops stronger resistance to external tidal influence. The outer regions are then the most prone to host a starburst activity triggered by the interaction-driven compression, as shown in non-cosmological simulations at higher resolution. This corresponds to the off-center starbursts observed in many interacting systems \citep{Whitmore1995, Smith2008, Hancock2009, Peterson2009, Elmegreen2016}. The middle panel of \fig{maps} shows \vintergatan and the progenitor of the last major merger (seen edge-on at the position ($12 \kpc, -17 \kpc$)) after the first pericenter passage and about $200 \Myr$ before their coalescence. In between the two galaxies, large volumes of intergalactic gas lie in the tidal compression created by the overlap of the extended galactic potentials. As noted before, this type of local compression is short-lived, but can be significant enough to ignite star formation (see \paperiii on the onset of star formation in the outer, tilting disc).

Finally, in the \emph{secular} phase, the situation becomes similar to that in the inner galaxy of the \emph{starburst} phase, but now across the entire galaxy. The absence of major mergers implies the lack of large-scale interaction-triggered tidal compression. The compression is limited to punctual and localized events linked with the accretion of satellites, and the intrinsic shape of the galactic potential. Only a small fraction of the gas mass is concerned, and the compression remains marginally weak (\fig{avel1}).

\begin{figure}
\centering
\includegraphics{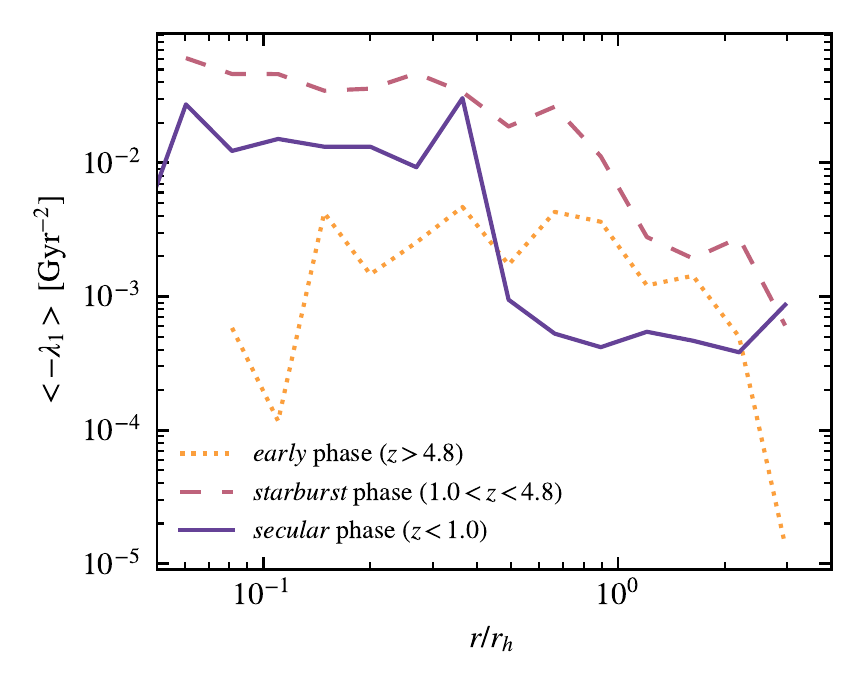}
\caption{Mass-weighted average strength of the compressive tidal field (showed as the log of $-\lambda_1$) in radial bins normalized to the stellar half-mass radius, and stacked from each snapshot of the 3 epochs.}
\label{fig:radiusstack}
\end{figure}

\fig{radiusstack} summarizes the evolution of the intensity and spatial distribution of compression. It shows the radial profile of the mass-weighted average $\lambda_1$ (in compressive regions only). To compensate for the inside-out growth of the galaxy, the profiles are normalized to the stellar half-mass radius. The profiles of each snapshot of the three phases are stacked to illustrate general trends, but at the expense of smoothing out punctual and localized events. The results discussed above can be transposed in this figure: a weak compression in the \emph{early} and \emph{secular} phases, and a stronger effect during the \emph{starburst} epoch. The effects of nearby galaxies inducing compression in the outer galaxy is also visible as radial extents of compression in the \emph{early} and \emph{starburst} phases. This figure illustrates again that off-centered tidal compression appears only in the presence of interactions (\emph{early} and \emph{starburst}), but that it becomes strong enough to trigger off-centered starbursts only once the disc is in place (\emph{starburst}).

%%%%%%%%%%%%%%%%
\subsection{Excess of dense gas}

\begin{figure}
\centering
\includegraphics{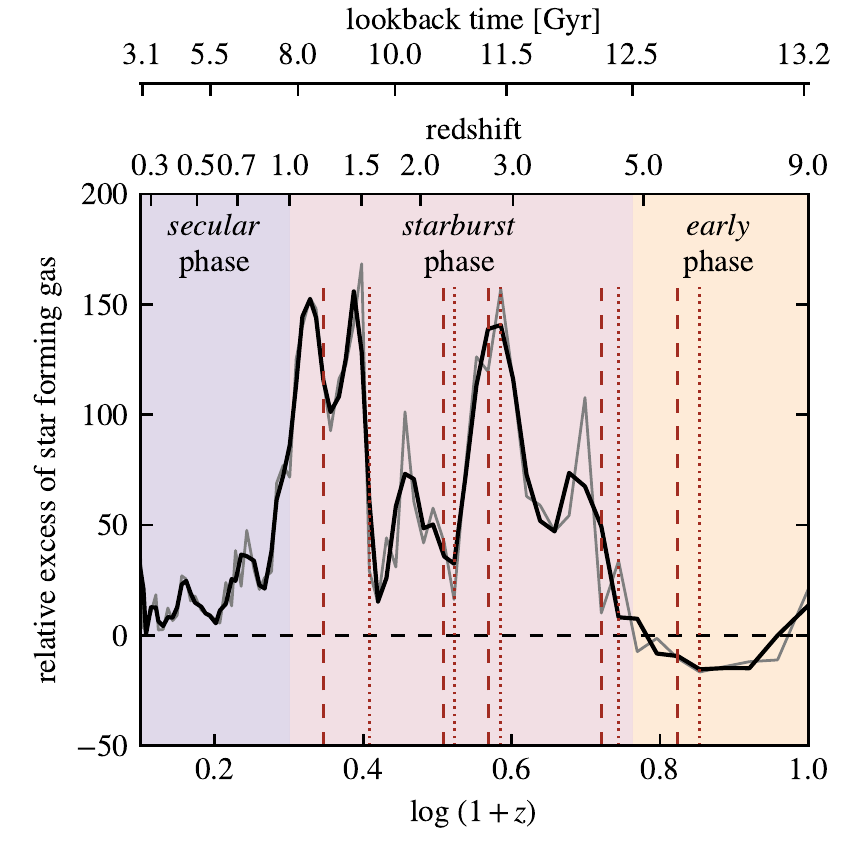}
\caption{Evolution of the excess of star forming gas, relative to the density distribution on the last snapshot (see text). As above, the raw data (grey) is smoothed using the Savitzky-Golay algorithm (black line), the shaded areas marks the three phases of stars formation, and the vertical lines indicate the pericenter passages and coalescences of major mergers. Strong excesses of dense gas are associated with major mergers, but only during the \emph{starburst} phase.}
\label{fig:excess}
\end{figure}

An increase of the SFR can originate from three types of alterations of the distribution of gas densities (and their combination): (\emph{i}) a global increase of the density, i.e. a shift of the entire density probability distribution function (PDF) towards the high density end. This is not possible at galactic scale, as it would mean the absence of low densities and thus the deconnection of the galaxy from its diffuse environment. (\emph{ii}) An overall increase of the gas mass, keeping the density PDF unchanged. (\emph{iii}) The formation of an excess of dense gas, thus distorting the PDF. (One can take the convenient but approximative shortcut of saying that the molecular fraction is increased in the latter case only.) The second case can be seen as a scaling transformation, as it happens during the adiabatic evolution of a galaxy: the amount of star forming gas changes self-similarly. In this case, the unchanged PDF indicates that the physical processes setting the density contrasts in the interstellar medium (e.g. turbulence, shocks, compression) are unchanged. This usually leads to a log-normal shape\footnote{For simplicity, we ignore here the power-law tail at high density observed and simulated in PDFs at cloud scales \citep[e.g.][]{Klessen2000, Elmegreen2011}.} of the density PDF \citep{Vazquez1994, Nordlund1999}. Analytical models show that increasing the gas mass in such a way moves the galaxy along the Kennicutt-Schmidt relation \citep{Renaud2012}. In the latter case however, the excess of dense gas reflects a modification of the mechanisms driving the density contrasts: the PDF is no longer log-normal. Because of the non-linearity of star forming laws\footnote{For instance, a widely accepted star formation law connects the local density of SFR $\rho_{\rm SFR}$ with the local density of gas $\rho$, normalized by the free-fall time $\tff$, i.e. $\rho_{\rm SFR} \propto \rho/\tff \propto \rho^{1.5}$ \citep{Elmegreen2002}.}, an excess of dense gas implies that the galaxy deviates from the classical Kennicutt-Schmidt relation. With a toy model to describe such an excess, \citet{Renaud2012} showed that such a galaxy can reach the starburst regime in the Kennicutt-Schmidt plane \citep{Daddi2010b}. This has later been confirmed with simulations of mergers \citep[e.g.][]{Renaud2014b, Renaud2019b}. The excess of dense gas is thus essential to reach an accelerated star formation (short depletion time), instead of a mere scaling up of the SFR as in the first case. This nuance tells apart starbursting galaxies from galaxies with a high SFR.

In the figure 4 of \paperiv, we show that the \emph{average} gas density PDFs of the \emph{starburst} phase (stacked from all the snapshots of this phase) exhibits an excess of the star forming gas with respect to the other two phases. This indicates a change in the physical process(es) setting the gas density distribution during this phase. \fig{excess} complements this analysis by highlighting that the excesses coincide with epoch of interactions, and thus of tidal compression. To compute the relative excess of dense gas shown in this figure, we first measure the PDF for each snapshot as the histogram of mass-weighted densities, and normalized by the total gas mass (as in \paperiv). Then, this histogram is normalized to its value at $100 \cc$, corresponding to the density threshold for star formation adopted in \vintergatan. We compute the relative difference between this distribution and that of the final snapshot, only considering the star forming gas ($> 100 \cc$). Finally, we compute the sum of the histogram bins. In other words, the dimensionless quantity shown in \fig{excess} can be interpreted as the amount of star forming gas in excess relative to what would be found in a secularily-evolving galaxy of the same gas mass. (Following the approximative shortcut proposed earlier, this would be the molecular fraction, relative to that of a non-starbursting disc galaxy.)

\fig{excess} shows strong excesses of dense gas at the epochs of tidal compression and short depletion times. It also shows a deficit of dense gas (with respect to a ``normal'' galaxy portraited by the last snapshot) during the \emph{early} phase, due to the shallow galactic potential and the inefficiency of interactions in triggering strong compression. This result expands the conclusion found in non-cosmological simulations of mergers where interaction-induced tidal compression (and subsequently turbulent compression) induce similar excess of dense gas \citep{Renaud2014b}. The picture holds in cosmological context, i.e. with the addition of gas accretion and repeated mergers.

%%%%%%%%%%%%%%%%
\section{Discussion}

%%%%%%%%%%%%%%%%
\subsection{Merger-induced starbursts in gas-rich disc galaxies}

The typical morphology of star forming disc galaxies at high redshift ($z \approx 1\mh 3$) is dominated by a handful of massive gas clumps \citep{Cowie1995, Elmegreen2007, Forster2009, Wuyts2012, Zanella2015, Guo2018}. Violent disc instabilities due to high gas fractions drive these differences with respect to contemporary discs, in terms of morphology, density distribution, and velocity dispersion \citep{Elmegreen2008, Dekel2009, Agertz2009, Inoue2016, Renaud2021c, vanDonkelaar2022, Ejdetjarn2022}. In galaxy simulations run in isolation, the instability regime leading to massive clump formation, instead of e.g. spiral structures, is found for gas fractions $\gtrsim 20\%$ \citep{Renaud2021c}.

\citet{Perret2014} and \citet{Fensch2017} presented a series of non-cosmological runs of mergers of clumpy galaxies, and reported very mild enhancements of their SFRs at the times of interactions, much weaker than for a merger of galaxies with lower gas fraction on the same orbit. After having ruled out feedback as a possible cause of stabilisation of the dense clumps, \citet{Fensch2017} proposed that the intrinsically strong turbulence of gas-rich discs saturates during interactions, thus not allowing the boost of density contrast needed to accelerate star formation, in particular in off-center regions. However, the physical origin of such a saturation and what would set its level remain unknown.

The non-cosmological mergers of \citet{Moreno2021} involve galaxies of comparable stellar mass as \vintergatan at $z=1$ (a few $10^{10}\Msun$), but with extremely high gas fractions: $\approx 70\%$ and $85\%$ (compared to $\approx 20\%$ in \vintergatan at $z=1$, which is in line with observations, see e.g. \citealt{Zanella2018} and references therein). Despite their high gas fraction, the discs of \citet{Moreno2021} do not have a clumpy morphology, and yield merger-induced enhancements of the SFR. The reason for the absence of massive clumps in such gas-rich galaxies is unclear, but is likely related to different implementations of feedback between \citet[which uses the FIRE-2 implementation of \citealt{Hopkins2018}]{Moreno2021} and the runs showing clumpy morphologies \citep[see e.g.][and their respective feedback recipes based on \citealt{Renaud2013b} and \citealt{Agertz2013, Agertz2015}]{Bournaud2015, Fensch2017, Renaud2021c}. Exploring this question is out of the scope of the present paper.

Although \vintergatan shows some clumps at $z\approx 1\mh 3$, the galaxy does not harbor a morphology dominated by an handful of massive clumps, as expected from simulations of comparable systems and resolution, with the same sub-grid recipes, but run without cosmological context \citep{Renaud2021c}. The reason for the absence of clumpy morphology is thus likely the intense merger bombardment \vintergatan undergoes at $z \gtrsim 1$, which repeatedly destroys most substructures in the disc.

Therefore, the merger-driven starburst activity in gas-rich discs noted in \citet{Moreno2021} and in \vintergatan, in contrast with its absence in \citet{Perret2014} and \citet{Fensch2017}, suggests that the inefficiency of mergers in triggering starbursts is rather connected with the morphology of the galaxy than with its gas fraction. Cosmological zoom-in simulations with different gas-consumption histories, different merger histories, and different regulations of star formation are required to check this hypothesis. If proven valid, this conclusion could then be extended to very high redshifts when the absence of a disc morphology also prevents the onset of merger-triggered starbursts (in our \emph{early} phase).

%%%%%%%%%%%%%%%%
\subsection{Starbursts cannot be identified from their sSFR}

\begin{figure}
\centering
\includegraphics{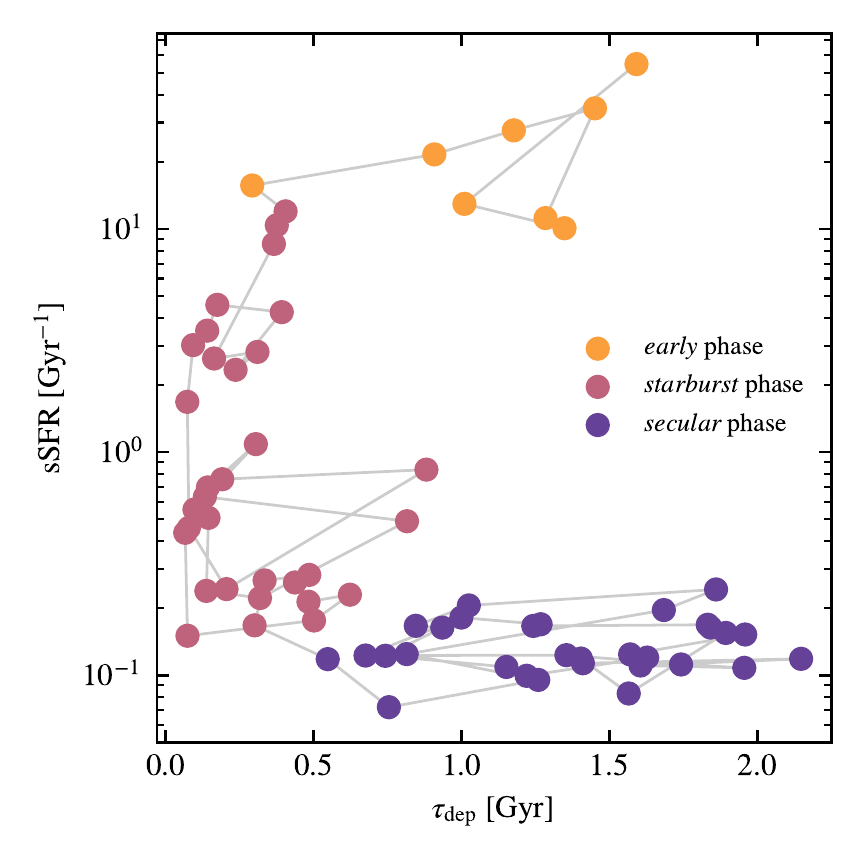}
\caption{Evolution of \vintergatan in the depletion time - sSFR plane. Points mark snapshots in the simulation. The three phases defined above correspond to clearly distinct regimes in this plane. The absence of correlation across the phases indicates that the outliers above the main sequence cannot correspond to starbursting galaxies (i.e. short \tdep) at all redshifts.}
\label{fig:ssfr}
\end{figure}

Using the simulation, we can directly compare the diagnostics from the two definitions of starburst presented in the introduction: short depletion time, and offset of sSFR with respect to the main sequence. First, \fig{ssfr} shows the evolution of \vintergatan in the \tdep - sSFR plane. The transitions between our three phases mark remarkably clear changes in the relation between the sSFR and the depletion time. While correlations between the two quantities could be found within each individual phase (although with highly different scatters), none exist across the entire evolution of the galaxy.

These variations originate from the definitions of the two quantities. The sSFR involves the stellar mass of the galaxy, i.e. the outcome of its past star formation activity, while the depletion time can be seen instead as an instantaneous indicator. Therefore, depending on how fast a galaxy has acquired its stellar component along its \emph{past} evolution, it could reach a high sSFR (possibly even enough to become an outlier above the main sequence) independently of how fast it \emph{currently} converts its gas into stars. (\app{ms} discusses the position of \vintergatan with respect to the main sequence.) In \vintergatan, the high sSFR of the \emph{early} phase merely reflects the very high gas fraction of the young galaxy, i.e. the small importance of its stellar component. Then, in the \emph{starburst} phase, the increase of the stellar mass decouples from that of the gas, due to the rapid consumption of the gas reservoir and the launch of outflows by the starburst events (see \paperii). It is thus not surprising that starburst events yield very different signatures in sSFR between the two phases.

\begin{figure}
\centering
\includegraphics{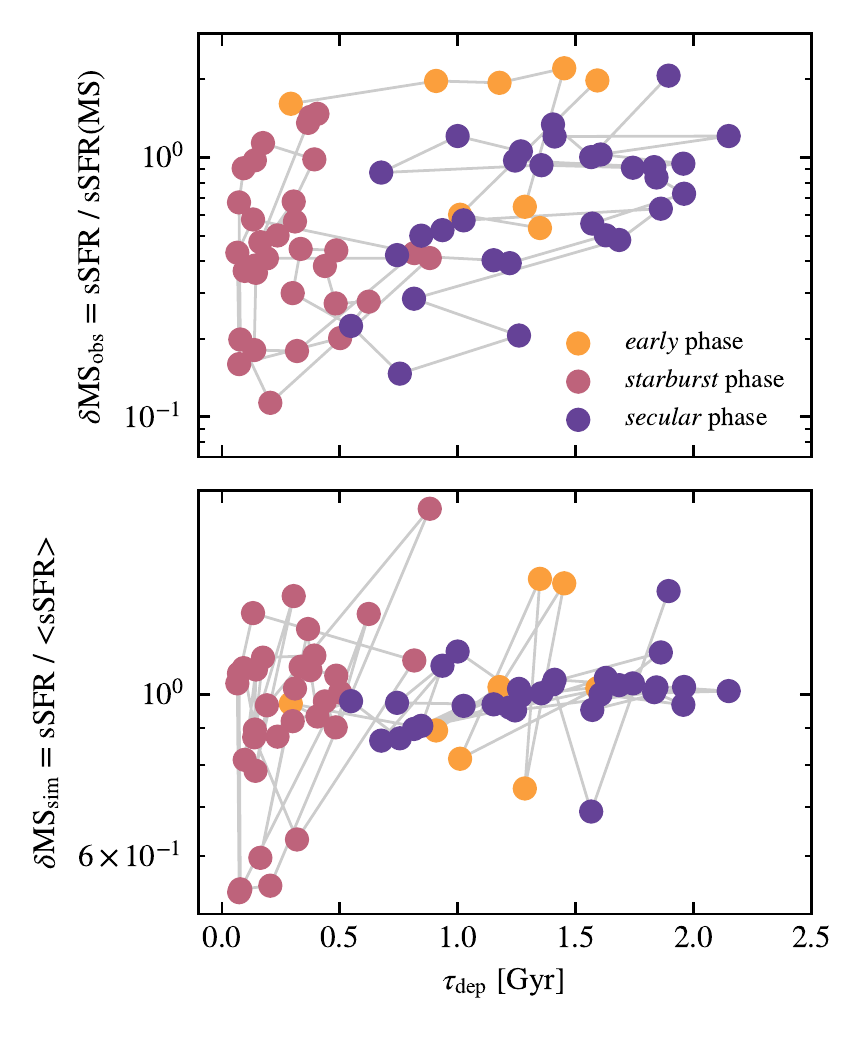}
\caption{Comparison of the depletion time with the offset of sSFR above the main sequence. The main sequence is defined using the relation from \citet{Tacconi2018} (top), and by smoothing the sSFR of \vintergatan over time (bottom).}
\label{fig:ssfr_tdep}
\end{figure}

Finally, \fig{ssfr_tdep} compares the two quantities commonly used in the literature to qualify starbursts: the depletion time and the offset with respect to the main sequence. In the top panel, we use the main sequence definition of \citet{Tacconi2018}, as in \fig{ms}. In the bottom panel, the offset is computed by normalizing the sSFR to its evolution smoothed using a Savitzky-Golay algorithm to eliminate the rapid variations. In both cases, the depletion time does not correlate with the offset above the main sequence. We note in particular a large scatter in the offset at short depletion times. This shows that, independently of the exact definition and normalization of the main sequence, outliers above the main sequence do not necessarily correspond to starbursts.

The degeneracy of the distributions in \fig[s]{ssfr} and \ref{fig:ssfr_tdep} demonstrates that even a redshift-dependent re-normalization of the main sequence (as commonly advocated, see \citealt{Pearson2018} and references therein) would not reconcile the definition of starburst based on a higher-than-average sSFR at a given stellar mass and redshift, with the starburst definition of fast star formation (i.e. short \tdep). Hence, independently of how the main sequence evolves, the classification of a galaxy within or above the main sequence does not necessarily corresponds to its ``normal'' or starburst nature. This confirms the observational results of \citet{Gomez2022} and \citet{Zavala2022} stating that starburst galaxies could lie within the scatter of the main sequence. Furthermore, it suggests that the bimodality in the distribution of sSFRs of star forming galaxies \citep[e.g.][]{Sargent2014, Rinaldi2022} is not, or at least not solely, explained by starburst events as rapid transitions above the main sequence. The divergence between the merger rate and the number of outliers above the main sequence \citep{Schreiber2015, Pearson2019} could thus be explained by a complex relation between outliers and starbursts, which depends on the phase of star formation.

%%%%%%%%%%%%
\section{Conclusion}

Using the \vintergatan cosmological zoom-in simulation of a Milky Way-like galaxy, we present a strong correspondence between the starburst activity and the tidal compression. Our main results are:
\begin{itemize}
\item Once the galactic disc is in place and the dynamics are well-ordered, galaxy interactions induce strong tidal compression.
\item This compression leads to an excess of dense gas, and a drop of the depletion time during the corresponding boosts of the SFR.
\item Strong compression can be found at large galacto-centric radii.
\item This mechanism is not active at high redshift before the galactic disc is in place, and neither after the last major merger because of the absence of external stimuli.
\item The correspondence between starbursts episodes (short depletion times) and high sSFR does not exist at all redshifts. This suggests that the distinction between galaxies on the main sequence and outliers above it does not necessarily correspond to the difference between normal star forming galaxies and starbursts.
\end{itemize}

One could question whether compression is the cause or the consequence of starbursts. For instance, supernova feedback locally increases the compressive mode of turbulence \citep{Grisdale2017}. Thus, turbulent compression increases after an episode of intense star formation. However, the properties of the tidal field result from the density distribution of all mass constituents of the galaxy, including stars and dark matter which are (directly) insensitive to feedback. Therefore, feedback is unlikely responsible for large-scale changes in the tidal field and large-scale compression\footnote{We note however that the repeated and localized injection of energy and momentum by stellar feedback is invoked in the cusp-to-core transformation of dwarf galaxies \citep[see][and references therein]{Read2016}. Interestingly, by creating a core, this transformation induces the onset of compressive tides at the center of these dwarfs. It is then likely that this favors further star formation, and thus sustains the source of feedback necessary to maintain the core profile.}. Furthermore, simulations with a higher time sampling have shown that, during an interaction, tidal compression occurs a few $10 \Myr$ \emph{before} the enhancement of star formation, and thus even longer before the associated feedback \citep{Renaud2014b, Li2022}. This causality argument demonstrates that starbursts are a consequence and not a cause of the elevation of the velocity dispersion induced by tidal and turbulent compression.

Compressive tides provide special physical conditions for the interstellar medium and star formation: mainly, the accumulation of dense gas over large volumes and a fast and enhanced star formation activity. Their importance in mergers suggests that they play a role in the formation and/or hierarchical assembly of young massive star clusters ($\gtrsim 10^4 \Msun$) commonly observed in interacting systems \citep[e.g.][]{Whitmore1995, Zepf1999, Forbes2000, Reines2008, Whitmore2014, Miah2015, Adamo2020}. While limited by resolution and missing collisional dynamics, several simulations of galaxy mergers have shown indeed that the initial cluster mass function could be significantly skewed toward high masses in starbursting mergers (\citealt{Renaud2015, Maji2017, Lahen2019b, Moreno2019, Li2022}). Tidal compression participates in this process, in particular in the off-center regions where other triggers of starburst like nuclear inflows are absent (see \citealt{Renaud2019b}). A parallel between young massive clusters in local mergers and globular clusters at the epoch of their formation is often made in the literature. However, if compression is indeed a formation channel of massive clusters, our results indicate that it could be too weak to play this role in the early Universe, when the potential of galactic discs is not yet in place. Therefore, the physical conditions of local mergers cannot be found at the epoch of globular cluster formation. This illustrates the vital need for star (cluster) formation models based on first principles, and not solely constrained by observations limited to the local Universe.

%%%%%%%%%%%%%%%%%%%%%%%%%%%%%%%%%%%%%%%%%%%%%%%%%%%%%%%%%%%%%%%%%%%%%%%%%%%%%%%%%
%%%%%%%%%%%%%%%%%%%%%%%%%%%%%%%%%%%%%%%%%%%%%%%%%%%%%%%%%%%%%%%%%%%%%%%%%%%%%%%%%
%%%%%%%%%%%%%%%%%%%%%%%%%%%%%%%%%%%%%%%%%%%%%%%%%%%%%%%%%%%%%%%%%%%%%%%%%%%%%%%%%
\section*{Acknowledgements}
We thank the referee and editor for their input, and David Elbaz for insightful discussions. We acknowledge support from the Knut and Alice Wallenberg Foundation, and from the Swedish Research Council (grant 2019-04659). 

\section*{Data availability}
The data underlying this article will be shared on reasonable request to the corresponding author.

\bibliographystyle{mnras}
\bibliography{biblio}

%%%%%%%%%%%%%%%%% APPENDICES %%%%%%%%%%%%%%%%%%%%%

\appendix

\section{Outliers above the main sequence}
\label{sec:ms}

\begin{figure}
\centering
\includegraphics{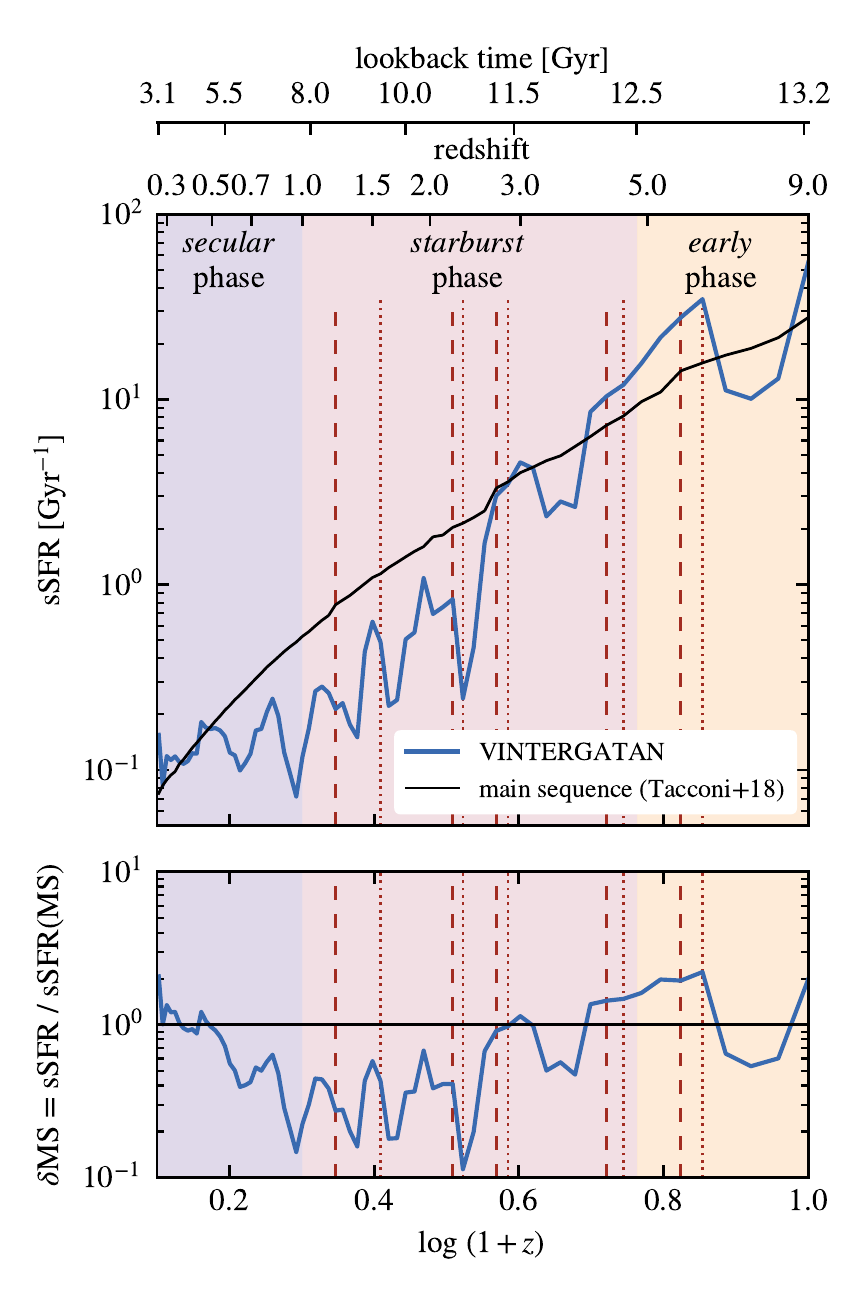}
\caption{Top: evolution of the sSFR of \vintergatan, compared to the expected value for the main sequence at same stellar mass (equation 1 of \citealt{Tacconi2018}, see also \citealt{Speagle2014}). Bottom: offset of \vintergatan with respect to this main sequence relation.}
\label{fig:ms}
\end{figure}

\fig{ms} shows the evolution of \vintergatan with respect to the main sequence computed from the relation of \citet[their equation 1, see also \citealt{Speagle2014}]{Tacconi2018}, and using the actual stellar mass from the simulation. While \vintergatan matches the observational relation at low and very high redshift, it significantly diverges from it between $z\approx 0.6$ and 3. This divergence does not match the epochs of peaks in the SFR, and even lasts for about 2 Gyr after the last major merger. It originates from the fact that \vintergatan acquires its stellar mass faster than the average galaxies observed on the main sequence, as also seen in other cosmological simulations \citep[e.g.][]{Vogelsberger2014, Sparre2015, Dubois2021}. This formation history implies that most of the starburst episodes of \vintergatan do not qualify as outliers above the observed main sequence.

The limited volume of our zoom-in simulation does not allow us to define a simulated main sequence based on a population of galaxies. However, we can invoke the ergodic principle to replace the average over a population with the average over time. We then define a proxy for the main sequence either by computing the moving time average of the sSFR, or by smoothing its time evolution to eliminate the peaks. Doing so makes to the peaks of sSFR correspond to offsets above this simulated main sequence by factors of $\approx 1.5\mh 2$. (This is largely insensitive to the details of the time averaging and smoothing procedures.)

\end{document}

%% file: header.tex
\usepackage{amssymb,latexsym,graphicx,natbib,eufrak,times,amsmath,xspace,ifthen}
\usepackage[normalem]{ulem}

\ifthenelse{ \draftversion > 0 }{
  \newcommand{\so}[1]{\color[rgb]{0.6,0,0.6}\sout{#1}\color{black}} % Strikethrough 
} {
  
  \newcommand{\so}[1]{}
}

\ifthenelse{\equal{\draftversion}{1}}{
  \usepackage{xcolor}
  \newcommand{\sep}[1]{\par\begin{color}[rgb]{0,0.4,0}\begin{center}\hrule\end{center}\end{color}\par} % SEPARATOR: green horizontal line
  \newcommand{\todo}[1]{\begin{color}{red}\ \ifthenelse{\equal{#1}{}} {$\bullet\bullet\bullet$} {$\bullet$\ #1 $\bullet$}\end{color}} % TODO: red
  \newcommand{\idea}[1]{\begin{color}[rgb]{0,0.4,0}\textit{#1}\end{color}} % IDEA: green
  \newcommand{\sk}[1]{\begin{color}[rgb]{0.6,0,0.6}#1\end{color}} % SKIP: purple
  \newcommand{\toc}{\par\begin{color}[rgb]{0.6,0,0.6}\begin{center}\hrule\vspace{0.5mm}\begingroup\small\let\cleardoublepage\relax\let\clearpage\relax\mytoc\endgroup\vspace{0.5mm}\hrule\end{center}\end{color}\par} % TOC

  }{
  \newsavebox{\trashcan}

  \newcommand{\sep}[1]{}
  \newcommand{\todo}[1]{}
  \newcommand{\idea}[1]{}
  \newcommand{\sk}[1]{}
  \newcommand{\toc}{}

  }
 \makeatletter\newcommand\mytoc{\@starttoc{toc}}\makeatother % set TOC style
\long\def\symbolfootnote[#1]#2{\begingroup%
\def\thefootnote{\fnsymbol{footnote}}\footnote[#1]{#2}\endgroup} 

\newcommand{\fig}[2][]{Figure#1~\ref{fig:#2}}

\newcommand{\sect}[2][]{Section#1~\ref{sec:#2}}
\newcommand{\app}[2][]{Appendix#1~\ref{sec:#2}}

\newcommand{\bb}[1]{\ifmmode \mbox{\boldmath $ #1$} \else  \mbox{\boldmath $#1$} \fi}

\newcommand{\mh}{\ensuremath{\textrm{\,--\,}}}    % long hyphen
\newcommand{\U}[1]{\ensuremath{\mathrm{~#1}}}     % units
\newcommand{\yr}{\U{yr}}
\newcommand{\Myr}{\U{Myr}}          \newcommand{\myr}{\Myr}
\newcommand{\Gyr}{\U{Gyr}}          
\newcommand{\pc}{\U{pc}}
\newcommand{\kpc}{\U{kpc}}
          
\newcommand{\Msun}{\U{M}_{\odot}}   
\newcommand{\Msunyr}{\Msun\yr^{-1}} \newcommand{\msunyr}{\Msunyr}
   
\newcommand{\cc}{\U{cm^{-3}}}

\newcommand{\tff}{\ensuremath{t_\mathrm{ff}}}     % free fall time
\newcommand{\aco}{\ensuremath{\alpha_\mathrm{CO}}\xspace}     % alpha_CO
\newcommand{\tdep}{\ensuremath{\tau_\mathrm{dep}}\xspace}        % t_dep
\newcommand{\ramses}{{\small RAMSES}\xspace}

\newcommand{\vintergatan}{{\small VINTERGATAN}\xspace}

\newcommand{\lund}{Department of Astronomy and Theoretical Physics, Lund Observatory, Box 43, SE-221 00 Lund, Sweden}